\documentclass[aps,prl,reprint,showpacs,floatfix]{revtex4-1}
\usepackage[latin1]{inputenc}
\usepackage{amsmath}
\usepackage{amsfonts}
\usepackage{amssymb}
\usepackage{lmodern,dsfont}
\usepackage{graphicx}
\usepackage{color}
\usepackage[]{natbib}

\newcommand{\pd}{\partial}
\newcommand{\be}{\begin{equation}}
\newcommand{\ee}{\end{equation}}

\begin{document}

\title{Anomalous spatial diffusion and multifractality in optical lattices}

\author{Andreas Dechant$^{1,2}$ and Eric Lutz$^{1,2}$}
\affiliation{$^1$Department of Physics, University of Augsburg, D-86135 Augsburg, Germany\\
$^2$Dahlem Center for Complex Quantum Systems, FU Berlin, D-14195 Berlin, Germany}

\begin{abstract}
Transport of cold atoms in shallow  optical lattices is characterized by slow, nonstationary momentum relaxation. We here develop a projector operator method able to derive in this case a generalized Smoluchowski equation for the position variable. We show that this explicitly non-Markovian equation can be written as a systematic expansion involving higher-order derivatives. We use the latter to compute arbitrary moments of the spatial distribution and analyze their multifractal properties.  
\end{abstract}

\maketitle

Slow relaxation is a hallmark of complexity. The absence of clear separations of time scales in complex systems
often leads to  nonexponential decay laws and anomalous diffusion properties \cite{shl93}. Prominent systems exhibiting this kind of strange kinetics are cold atoms in dissipative optical lattices. Optical lattices are highly tunable periodic standing-wave potentials created by counter-propagating laser beams \cite{gry01}. By decreasing the  depth of the potential, the atomic dynamics can be modified in a controlled manner from diffusive/ergodic to anomalous/nonergodic \cite{lut03}. In the semiclassical regime, in which emission and absorption of single photons only slightly perturb the atomic motion, the latter can be described by a phase-space equation with nonlinear friction \cite{cas91}. This Klein-Kramers equation has been successfully used in theoretical \cite{mar96, per00}, as well as in experimental investigations \cite{hod95,dou06,kat97}. In particular, the predicted   non-Gaussian momentum distribution \cite{dou06}, the divergence of its second moment in shallow potentials \cite{kat97}, and the spatial superdiffusion \cite{kat97,sag11} have been experimentally observed. However,  the general equation governing  spatial diffusion, required to describe these experiments, is missing. 

In the theory of stochastic processes, the equation describing the spatial motion is usually derived with the help of the standard procedure of adiabatic elimination of fast variables \cite{gar84}. This method exploits the rapid relaxation of the momentum to a stationary state to extract an equation of the Fokker-Planck type for the slow position variable out of the full phase-space equation. The obtained equation is Markovian and does not depend on the initial condition. For shallow optical potentials, by contrast,  the momentum decay is algebraically slow \cite{mar96} and the momentum distribution nonstationary \cite{kes10}---a well-defined relaxation time, hence, does  not exist. As a result, the standard elimination technique cannot be used. In the present paper, we develop a general projector operator method that is able to handle both slow momentum relaxation and nonstationarity. When applied to the optical lattice, we obtain a generalized non-Markovian Smoluchowski equation in the form of a systematic expansion involving higher order derivatives. Similar expansions, though Markovian and relying on stationarity, have been considered in the past \cite{ris79,doe87}, in particular in the context of the Rayleigh model of Brownian motion with  finite-time collisions \cite{kam61,ply08}. They have been shown to provide more accurate descriptions than second-order equations \cite{ris79,doe87}. Common to these studies is the presence of a small parameter which allows the truncation of  the expansion. Due to the inherent scale invariance of the atomic dynamics, such a small parameter does not exist  in the case of the optical lattice. In the following, we use the generalized Smoluchowski equation to determine the generic moments of the spatial distribution, and reveal their multifractal structure. We further support our analytical predictions with detailed numerical simulations.

\textit{Projector operator method.} The Klein-Kramers equation for the phase-space density $W(x,p,t)$ is of the form 
\begin{align}
\pd_t W(x,p,t) = (L_x + L_p) W(x,p,t), \label{1}
\end{align}
where the linear operators $L_x$ and $L_p$ are given by
\begin{align}
L_x = -p \pd_x \  \quad\mbox{and} \quad L_p = \pd_p\left(-F(p) + D(p) \pd_p \right). \label{2}
\end{align}
The semiclassical momentum-dependent  drift and diffusion coefficients for  atoms in the optical lattice  are \cite{cas91,mar96,per00,hod95},
\begin{align}
F(p) = - \frac{\gamma p}{1 + \left(p/p_c\right)^2}, \quad D(p) = D_0 + \frac{D_1}{1+\left(p/p_c\right)^2} \label{2a},
\end{align}
where the friction coefficient $\gamma$, the diffusion constants $D_0$ and $D_1$, and the capture momentum $p_c$ can be directly computed from the microscopic Hamiltonian of the problem.
Our goal is to derive the Smoluchowski  equation for the position distribution, $W_x(x,t)=\int dp \, W(x,p,t)$. 
To this end, we introduce the projection operator $P=P^2$ such that  $P f(p,x,t) = W_p(p,t^*) \int dp' \, f(x,p',t)$ for any integrable  function $f(x,p,t)$. Here  $W_p(p,t^*)$ denotes the nonstationary solution of the momentum equation at time $t^*$ \cite{kes10}, which we assume to be large, but finite. In the quasistationary limit, $\tau = t-t^* \ll t^*$, we have  $L_p W_p(p,t^*)\simeq 0$. We check numerically the validity of this approximation below. The function $W_p(p,t^*)$ is  an even function of $p$ for odd friction  and even diffusion coefficients, like in Eq.~\eqref{2a}, so that $P L_x P = 0$. We define $A(x,p,\tau) = P W(x,p,\tau)$ and $B(x,p,\tau) = (1-P) W(x,p,\tau)$. The two functions $A$  and $B$ respectively
belong to the subspace of relevant and irrelevant variables.  The elimination method consists of
 two  steps \cite{gar84}. First, we apply the operators $P$ and $1 - P$ to Eq.~\eqref{1} to obtain two coupled equations
for $A$ and $B$. We then formally solve the equation for $B$ and substitute the result in the equation for $A$, thus  eliminating  the irrelevant variables. This procedure is most
conveniently carried out in Laplace space. Using the notation $\tilde{f}(x,p,s) = \int_{0}^{\infty} d\tau e^{-s \tau} f(x,p,\tau)$, we obtain the following  equations for the Laplace transforms
of $A$ and $B$:  
\begin{align}
s \tilde{A}(s) - A(0) &= P L_x \tilde{B}(s), \\
s \tilde{B}(s) - B(0) &= (L_x+L_p)  \tilde{A}(s) 
 + [L_p + (1-P)L_x] \tilde{B}(s). \nonumber \label{5}
\end{align}
Solving the second equation for $\tilde{B}(x,p,s)$, we find
\be
s  \tilde{A}(s) - A(0) 
=  P L_x H^{-1} [L_x  \tilde{A}(s) +L_p  \tilde{A}(s)+ B(0)], \label{6}
\ee
where we have defined  $H = s-L_p - (1-P) L_x $. Equation \eqref{6} is exact. For  fast momentum decay, it is customary to approximate $H^{-1}\simeq - L_p^{-1}$ \cite{gar84}. We here expand $H^{-1}$ with respect to the coupling term $(1-P)L_x$  \cite{com},
\begin{align}
H^{-1} = \sum_{n=0}^{\infty} [s-L_p]^{-(n+1)} (1-P)^n L_x^n, \label{7}
\end{align}
and use the transformation,
\begin{align}
[s-L_p]^{-(n+1)} &=  \left[ \int_{0}^{\infty} d\tau \, e^{(L_p-s)\tau}\right]^{n+1} \nonumber \\
&=\frac{(-1)^n}{n!} \int_{0}^{\infty} d\tau \, \tau^n e^{(L_p-s)\tau}. \label{8}
\end{align}
The integral over $\tau$ is finite, since  stability  requires $L_p$ to have nonpositive eigenvalues. Further noting that  $(1-P)^n = (1-P)$ ($n>0$), we can write Eq.~\eqref{7} as, 
\begin{align}
H^{-1} = & \int_{0}^{\infty} d\tau \, e^{(L_p-s)\tau} \nonumber \\
&+ \sum_{n=1}^{\infty} \frac{(-1)^n}{n!} \int_{0}^{\infty} d\tau \, \tau^n e^{(L_p-s)\tau} (1-P) L_x^n. \label{9}
\end{align}
For each value of $n$, we have to evaluate the three terms on the right-hand side of Eq.~\eqref{6}. The first term is 
\begin{align}
P &L_x \frac{(-1)^n}{n!} \int_{0}^{\infty} d\tau \, \tau^n e^{(L_p-s)\tau} (1-P) L_x^{n+1} \tilde{A}(s) \nonumber \\
&= \frac{1}{n!} \pd_ x^{n+2} \tilde{W}_x(x,s) W_p(p,t^*) \int d\tau \, \tau^n e^{-s\tau} \nonumber \\
&\times \int dp' p'^{n+1} W_p(p',t^*) E_p(p',\tau), \label{12}
\end{align}
where we have used the fact that $e^{L_p \tau} p^{n+1} W_p(p,t^*)$ is the solution of the  momentum equation with initial condition $f(p) = p^{n+1} W_p(p,t^*)$. We have accordingly, $e^{L_p \tau} p^{n+1} W_p(p,t^*) = \int dp' W_p(p,\tau \vert p',0) p'^{n+1} W_p(p',t^*)$ with the conditional probability density $W_p(p,\tau \vert p',0)$.
The second term  vanishes for stationary processes. For nonstationary processes, it is of order $\tau/t^*$, and can hence be neglected in the limit $\tau \ll t^*$. The third term can be set to zero by further assuming that  the joint density factorizes at $\tau =0$, $W(x,p,0)=W_p(p,t^*)W_x(x,0)$. Since $E_p(p,\tau)= \int dp' p' W_p(p',\tau|p,0)$ is an odd function of $p$, Eq.~\eqref{12} is only non-zero for even $n$. Combining Eqs.~\eqref{6}-\eqref{12} and going back to time space, we eventually  obtain the generalized, non-Markovian Smoluchowski equation,
\be
\pd_\tau W_x(x,\tau) 
 = \sum_{n=2}^{\infty} \int d\tau'  C_{n}(\tau-\tau',t^*) \pd_x^{n} W_x(x,\tau'), \label{17}
\ee
with the generalized correlation functions,
\begin{align}
C_{n}(\tau,t^*) = \frac{\tau^{n-2}}{(n-2)!} \int dp \, p^{n-1} E_p(p,\tau) W_{p}(p,t^*). \label{18} 
\end{align}
For  $n=2$, Eq.~\eqref{18} gives the two-time momentum correlation function, $C_2(\tau,t^*)=\langle p(\tau + t^*)p(t^*)\rangle$. We stress that due to the nonstationarity of the momentum process, the correlation functions $C_{n}$ depend explicitly on  $t^*$. Furthermore, Eq.~\eqref{17} reduces to the usual stationary, Markovian, second-order Smoluchowski equation  in the long-time limit, $\tau \gg 1/\gamma$,  when the mean momentum  relaxes exponentially fast, $E_p(p,\tau)\sim e^{-\gamma \tau}$. In this case,  the  spatial diffusion coefficient is  $D_x = \int_{0}^{\infty} d\tau \, C_2(\tau)$ \cite{gar84}. For  algebraically slow momentum relaxation with no characteristic decay time,   the nonstationary, non-Markovian nature of Eq.~\eqref{17} will persist, even at long times, and the expansion cannot be truncated at the second order (see  below). We can next derive an  equation for the moments of the spatial distribution  by multiplying Eq.~\eqref{17} by $x^{m}$ and then integrating over $x$. After integrating  the $n$th term by parts $n$ times, we find,
\begin{align}
\label{18a}
\pd_\tau &\langle x^{m}(\tau) \rangle \\
&  =\sum_{n=2}^{m} \frac{m!}{(m-n-2)!} \int_{0}^{\tau} \!d\tau' C_{n}(\tau-\tau',t^*) \langle x^{m-n-2}(\tau') \rangle .  \nonumber 
\end{align}
The $m$th moment  is hence fully determined by the first $(m-1)$ moments and the correlation function \eqref{18}.

\textit{Application to optical lattices.} Let us now apply the above formalism to the problem of  transport of cold atoms in shallow dissipative optical lattices. 
For the drift and diffusion coefficients \eqref{2a}, the asymptotic long-time behavior of the mean momentum $E_p(p,t)$ is \cite{dec11},
\begin{align}
\label{14}
E_p(p,\tau) &\simeq \frac{\sqrt{\pi}}{2 \Gamma(\alpha+1)}(4 D_0 \tau)^{\frac{1}{2}-\alpha} p^{2 \alpha} e^{-\frac{p^2}{4 D_0 \tau}} \nonumber \\
& \quad \quad \times M \left(\frac{3}{2},\alpha + 1, \frac{p^2}{4 D_0 \tau} \right),
\end{align}
where $M(a,b,z)$ is the confluent hypergeometric function, $\Gamma(z)$ the Gamma function, and $\alpha = \gamma/(2 D_0) + 1/2$. For simplicity, we here set $D_1=0$ and $p_c=1$ (we will treat the general case  elsewhere \cite{dec12}). In addition, the nonstationary, finite-time momentum distribution $W_{p}(p,t^*)$ (the infinite covariant density) reads  \cite{kes10},
\begin{align}
\label{15}
W_p(p,t^*) \simeq \left\lbrace \begin{array}{ll}
\frac{1}{Z \Gamma(\alpha)}  p^{1-2\alpha}\Gamma\left(\alpha, \frac{p^2}{4 D_0 t^*} \right)  \hspace{5mm}\qquad (\alpha > 1) \\[2 ex]
\frac{1}{\Gamma(1-\alpha)}(4 D_0 t^*)^{\alpha - 1}  p^{1-2\alpha} e^{\frac{p^2}{4 D_0 t^*}}\quad (\alpha < 1)
\end{array} \right.
\end{align}
with $\Gamma(\alpha,z)$  the lower incomplete Gamma function and $Z= \sqrt{\pi}\Gamma(\alpha-1) /\Gamma(\alpha-1/2)$ the partition function. We will use the above expressions to evaluate the asymptotic  behavior of the spatial moments \eqref{18a} to arbitrary order. 

\begin{figure}
\includegraphics[trim=30mm 45mm 20mm 15mm, clip, width=0.49\textwidth]{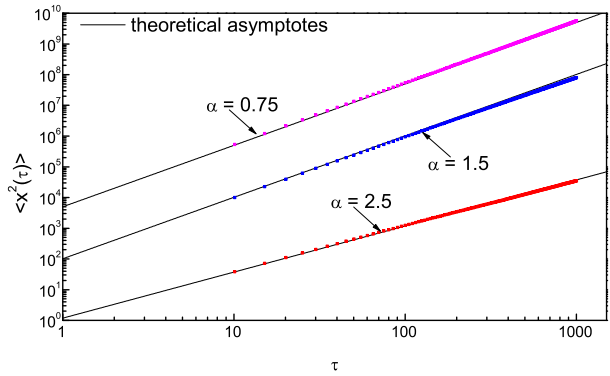}
\caption{(color online) Second moment of the position distribution, Eq.~\eqref{24a}, for three different values of the parameter $\alpha$ (black solid). The colored dots are the results of Langevin simulations. Parameters are  $D_0=1$ and $t^* = 5000$.}
\label{fig1}
\end{figure}

\begin{figure}
\includegraphics[trim=30mm 45mm 20mm 15mm, clip, width=0.49\textwidth]{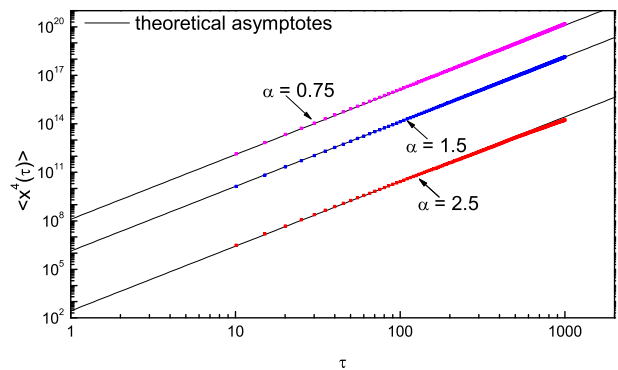}
\caption{(color online) Fourth moment of the position distribution, Eq.~\eqref{24b}, for three different values of the parameter $\alpha$ (black solid). The colored dots are the results of Langevin simulations. Parameters are  $D_0=1$ and $t^* = 5000$.}
\label{fig2}
\end{figure}

The coefficients $C_{n}(\tau,t^*)$ are directly related to the $n$th moment of the position distribution via Eq.~\eqref{18a}. Evaluating the asymptotic behavior of Eq.~\eqref{18} with the help of  Eqs.~\eqref{14} and \eqref{15}, we find  the general expressions,
\begin{align}
\langle &x^{n}(\tau)\rangle - \langle x^{n}(t^*)\rangle \simeq \hat{c}_{\alpha,n} \nonumber \\
&  \times \left\lbrace \begin{array}{ll}
\tau^{\frac{n}{2}} & \text{for} \quad \alpha > n + 1 \\[2 ex]
\tau^{\frac{3 n}{2}+1-\alpha} & \text{for} \quad \frac{n}{2}+1 < \alpha < n + 1 \\[2 ex]
{t^*}^{\frac{n}{2}+1-\alpha} \, \tau^{n} & \text{for} \quad 1 < \alpha < \frac{n}{2}+1 \\[2 ex]
{t^*}^{\frac{n}{2}} \, \tau^{n} & \text{for} \quad \alpha < 1 
\end{array} \right. \label{24}
\end{align}
Equation~\eqref{24} predicts normal behavior of the  $n$th moment for $\alpha > n + 1$, anomalous, subballistic superdiffusion for $n/2+1 < \alpha < n+1$, and quasiballistic superdiffusion for $\alpha < n/2+1$ and $\alpha < 1$. The asymptotic formulas of the coefficients $\hat{c}_{\alpha,n}$ can be  computed analytically for all $n$ using Eq.~\eqref{18a} \cite{dec12}.  We here only provide the two lower order terms. For $n=2$, we obtain,
\begin{align}
&\langle x^2 (\tau) \rangle - \langle x^2 (t^*) \rangle \nonumber \\
&\simeq \left\lbrace \begin{array}{ll}
\frac{\pi (4 D_0)^{2 - \alpha} \Gamma\left(\alpha - 2\right) \Gamma\left(3-\alpha\right)}{2 Z \, \Gamma\left(\alpha - \frac{1}{2}\right)^2 \Gamma(5-\alpha)} \tau^{4-\alpha} &\text{for} \quad 2 < \alpha <  3 \\[2 ex]
\frac{(4 D_0)^{2 - \alpha}}{Z \, \Gamma(\alpha) \left(2 - \alpha\right)} {t^*}^{2-\alpha} \tau^2 &\text{for} \quad 1 < \alpha < 2 \\[2 ex]
4 D_0 (1-\alpha) t^* \, \tau^2 &\text{for} \quad \alpha < 1 
\end{array} \right. \label{24a}
\end{align}while, for $n=4$, we find, 
\begin{align}
&\langle x^4 (\tau) \rangle - \langle x^4 (t^*) \rangle \nonumber \\
&\simeq \left\lbrace \begin{array}{ll}
\frac{2 \, (4 D_0)^{3 - \alpha}}{Z \, \Gamma(\alpha) \left(3-\alpha\right)} {t^*}^{3-\alpha} \tau^4 &\text{for} \quad 1 < \alpha < 3 \\[2 ex]
(4 D_0)^2 (1-\alpha) (2-\alpha) {t^*}^2 \, \tau^4 &\text{for} \quad \alpha < 1  \label{24b}
\end{array} \right.
\end{align}
The second moment has been measured in the regime $1<\alpha<3$ in Ref.~\cite{kat97}, and the transition from normal to superdiffusion has been observed. The case $\alpha<1$ has been examined  recently,  and ballistic behavior has been seen \cite{sag11}.
In Figs.~\ref{fig1} and \ref{fig2}, we compare the theoretical results, Eqs.~\eqref{24a} and \eqref{24b}, for the second and fourth moment with Langevin simulations of the full dynamics \eqref{1} (Euler method with time step 0.1 for $2.4\times10^6$ trajectories). The excellent agreement confirms the validity of the quasistationary aprroximation used in their derivation. In both cases, the prefactor depends explicitly on the initial time  $t^*$, indicating the nonstationarity of the process. Let us emphasize two important points. First, the second moment is often calculated as a double time integral of the two-time momentum correlation function \cite{dec11}. Trying to extend this approach to higher moments appears a formidable task since it involves the computation of $n$-time momentum correlation functions, contrary to the present method which only requires the generalized two-time correlation functions $C_n$. On the other hand, the $n$th moment of the equilibrium  distribution, $W_{\text{eq}}(p) = (1/Z)( 1 + p^2)^{(1/2-\alpha)}$, diverges when $n>2 \alpha-2$. Hence, had we used the latter in the definition of the projector operator $P$, as is done in the standard elimination procedure \cite{gar84}, the corresponding correlation functions $C_n$, Eq.~\eqref{18}, would not have been defined.  

An essential property of the generalized Smoluchowski equation \eqref{17} is that at least $n$ terms in the expansion are necessary to correctly compute the $n$th moment. For instance, terminating the expansion \eqref{17} at the second order  yields
 the  fractional diffusion equation  ($2 < \alpha < 3$),
\begin{align}
\frac{\partial}{\partial \tau} W_x(x,\tau) = c_{\alpha,2} \pd_{\tau}^{3-\alpha} \frac{\partial}{\partial x^2} W_x(x,\tau), \label{23}
\end{align}
where we have introduced  the Riemann--Liouville fractional derivative, $\pd_t^{-\nu} f(t) = 1/\Gamma(\nu) \int_0^t dt' (t-t')^{\nu-1} f(t')$, with $0\!<\!\nu\!<\!1$ \cite{sai97}. Equation~\eqref{23} describes a subballistic superdiffusive process. Its solution can be expresssed in terms of  a Fox function and satisfies the scaling form \cite{met00} 
\begin{align}
W_x(x,t) = t^{\frac{\alpha-4}{2}} g(\xi) \quad \text{with} \quad \xi = x t^\frac{\alpha-4}{2}. \label{23b}
\end{align}
The latter  implies the   scaling law for the moments:
\begin{align}
\langle \vert x \vert^q (\tau) \rangle \propto \tau^{\frac{4-\alpha}{2} q}. \label{23c}
\end{align}
Equation \eqref{23c}, however, only agrees with the general expression \eqref{24}  for the second moment. Higher-order moments require higher-order terms to be taken into account.  
These terms, in turn, will break the simple scaling of Eq.~\eqref{23}, leading to a complex multifractal structure.

\begin{figure}
\includegraphics[trim=30mm 45mm 20mm 15mm, clip, width=0.50\textwidth]{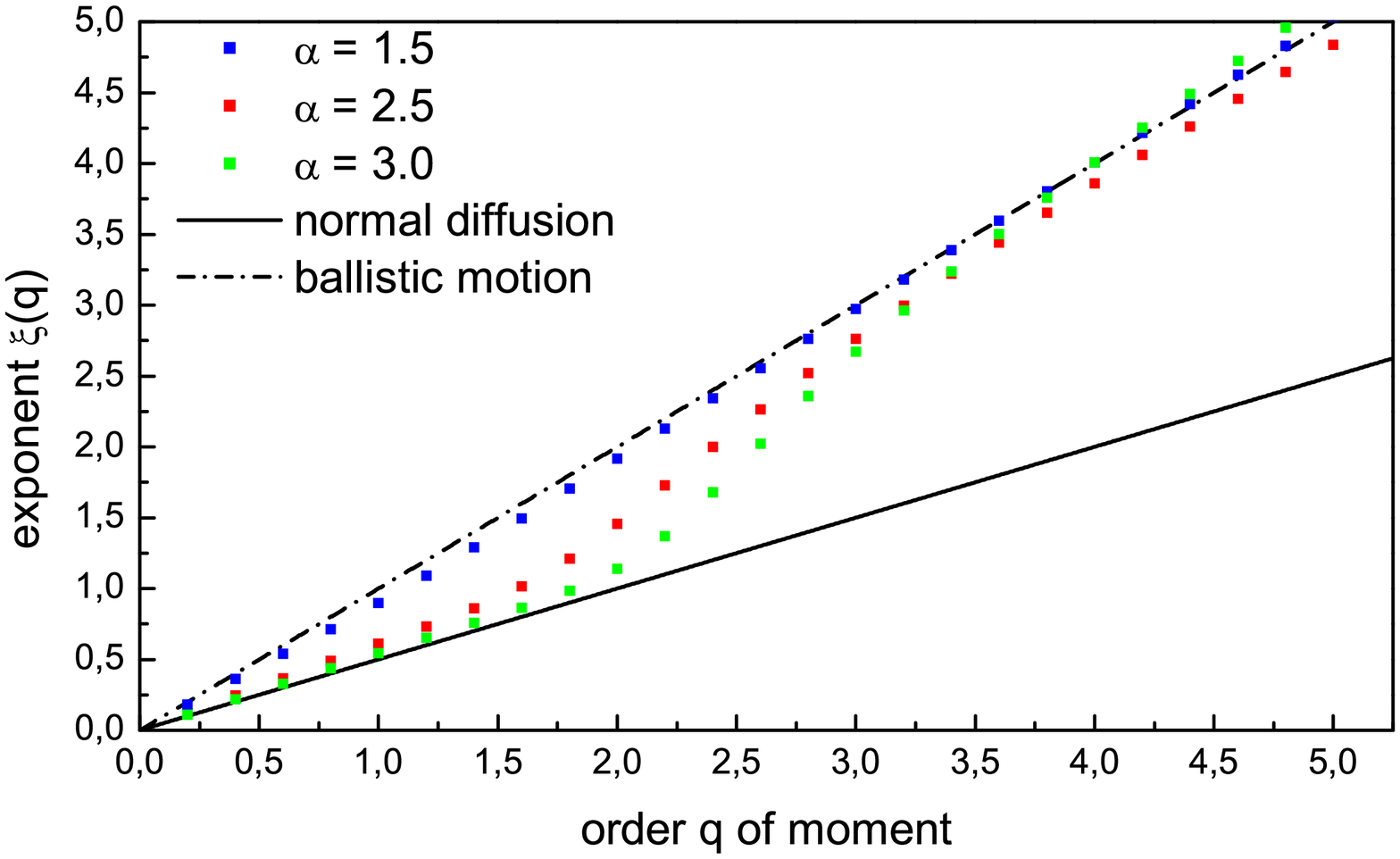}
\caption{(color online) Exponent $\xi(q)$ of the moments $\langle \vert x \vert^q(\tau) \rangle \propto \tau^{\xi(q)}$ as a function of the order $q$, for different values of $\alpha$.  The solid line corresponds to normal Gaussian diffusion ($\xi(q) = q/2$), whereas the dash-dotted line represents quasiballistic expansion ($\xi(q) = q$). The solid squares are obtained by fitting the long-time behavior of Langevin simulations. The multifractality of the process is clearly visible, as $\xi(q)$ is not a linear function of $q$. It interpolates between normal diffusion for  lower-order  and quasiballistic motion for  higher-order moments. Same parameters as in Fig.~(1).
}
\label{fig4}
\end{figure}

\textit{Multifractality.} The concept of multifractality quantifies the  variation of the scaling properties of a process \cite{sor06}. It often results from the overlapping of different scaling behaviors \cite{sta88}.
 Consider the $q$th moment \cite{bou00},
\begin{align}
\langle \vert x \vert^{q}(\tau) \rangle \propto \tau^{\xi(q)}. \label{27a}
\end{align}
A process is said to be monofractal, if the function $\xi(q)$ is linear, $\xi(q) = c \, q$ (normalization implies $\xi(0)=0$). For example, $c = 1/2$ for  Gaussian diffusion, while $c = 1$ for ballistic motion. For a multifractal process, $\xi(q)$ is a nonlinear function, indicating the presence of a continuous set of scaling exponents. 
Multifractality has been observed in many complex systems like, for example, turbulent and disordered mesoscopic systems (for a review see Ref.~\cite{pal87}).
For integer moments, $q=n$, we can determine the function $\xi(q)$ from  Eq.~\eqref{24}. We find,
\begin{align}
\xi(q) = \left\lbrace \begin{array}{ll}
\frac{q}{2} &\text{for} \quad q < \alpha - 1 \\[2 ex]
\frac{3 q}{2} + 1 - \alpha &\text{for} \quad  \alpha - 1 < q < 2 \alpha - 2\\[2 ex]
q &\text{for} \quad q > 2 \alpha - 2  \text{ or }  \alpha < 1.
\end{array} \right. \label{27}
\end{align}
Meanwhile, for noninteger moments, Fig.~\ref{fig4} shows that $\xi(q)$ is in general a nonlinear function of $q$. Hence, the exponent of the spatial $q$th moment exhibits a nontrivial  dependence on $q$ \textit{and} $\alpha$. For instance, for any given value of $\alpha$, the moments $\langle x^{q}(\tau) \rangle$ of order $q < \alpha - 1$ behave as for normal diffusion. This is directly related to the finiteness of the equilibrium moment $\langle p^{2q}\rangle$. By contrast, for $\alpha - 1 < q < 2\alpha - 2$, motion is subballistic superdiffusive, and even quasiballistic superdiffusive for $q > 2 \alpha - 2$. The latter property illustrates the non-Gaussian nature of the underlying process: Even if the second moment increases linearly with time ($\alpha > 3$), there always exist some higher-order moments which exhibit superdiffusive behavior. Thus, the second moment alone is not sufficient to fully capture the dynamics, as is typical for multifractal processes. For $\alpha < 1$, the process is monofractal.

\textit{Conclusion.}
We have developed a general projector operator method to derive a spatial Smoluchowski equation in situations where there is no time scale separation between position and momentum variables.  Our formalism is able to treat both algebraic momentum relaxation and nonstationary momentum dynamics. We have shown that spatial diffusion is described by a nonstationary and non-Markovian generalization of the Smoluchowski equation that can be written as a systematic expansion involving higher-order derivatives. We have applied this equation to the problem of transport of cold atoms in shallow dissipative optical lattices and obtained  explicit expressions for spatial moments of arbitrary order. We have further established their generic multifractal behavior. Because of the latter, a truncation of the expansion is not possible. As a consequence,  the lowest order term alone---given by a fractional diffusion equation---does not suffice to account for the complexity of the dynamics.

\acknowledgements{We thank E. Barkai, D. Kessler and F. Renzoni for discussions. This work was supported by the Emmy Noether Program of the DFG (contract No LU1382/1-1), the cluster of excellence Nanosystems Initiative Munich (NIM) and the Focus Area Nanoscale of the FU Berlin.}


\begin{thebibliography}{99}
\bibitem{com} We here omit the adjoint expression, which accounts for the noncommutability of $L_p$ and $L_x$,  since its inclusion does not change the final result.
\end{thebibliography}
\end{document}